\documentclass[]{aa}
\usepackage{graphicx}
\usepackage{txfonts}
\usepackage{longtable}
\usepackage{multirow}

\usepackage{natbib}
\bibpunct{(}{)}{;}{a}{}{,}
\begin{document}
   \title{The AMIGA sample of isolated galaxies}

   \subtitle{VII. Far-infrared and radio continuum study of nuclear activity\thanks{Full tables 1, 2, 4--6 are available in electronic form at http://www.iaa.csic.es/AMIGA.html and at the CDS web site http://cdsweb.u-strasbg.fr}}

   \author{
J. Sabater \inst{1}
\and
S. Leon\inst{2}
\and
L. Verdes-Montenegro\inst{1}
\and
U. Lisenfeld\inst{3}
\and
J. Sulentic\inst{4}
\and
S. Verley\inst{5}
          }
   \offprints{J. Sabater}

   \date{Received September 15, 1996; accepted March 16, 1997}
\institute{Instituto de Astrof\'{\i}sica de Andaluc\'{\i}a, CSIC,
           Apdo. 3004, 18080
           Granada, Spain\\
           \email{jsm@iaa.es}
\and
           Instituto de RadioAstronom\'{\i}a Milim\'etrica (IRAM), Granada, Spain
\and
          Dept. F\'{\i}sica Te\'orica y del Cosmos, Universidad de Granada,
          Spain
\and
           Department of Astronomy, Univ. of Alabama, Tuscaloosa, USA
\and
           Osservatorio Astrofisico di Arcetri, Istituto Nazionale di Astrofisica, Firenze, Italia
}
\date{Received ; accepted }


  \abstract
   {This paper is part of a series involving the AMIGA project (Analysis of the 
Interstellar Medium of Isolated GAlaxies). This project provides a 
statistically-significant sample of the most isolated galaxies in the northern 
sky.}
   {We present a study of the nuclear activity in a well-defined sample of the 
most isolated galaxies (total sample: n = 1050, complete subsample: n = 719) in 
the local Universe traced by their far-infrared (FIR) and radio continuum 
emission.}
   {We use the well-known radio continuum-FIR correlation to select 
radio-excess galaxies that are candidates to host an active galactic nucleus 
(AGN), as well as the FIR colours to find obscured AGN-candidates. We also 
used the existing information on nuclear activity in the V\'eron-Cetty 
catalogue and in the NASA Extragalactic Database.}
   {A final catalogue of AGN-candidate galaxies has been produced that will 
provide a baseline for studies on the dependence of activity on 
the environment. Our sample is mostly radio quiet, consistent with its high 
content of late-type galaxies. At most $\backsim 1.5\%$ 
of the galaxies show a radio excess 
with respect to the radio-FIR correlation, and this fraction even goes down to 
less than 0.8\% after rejection of back/foreground sources using FIRST. We find that 
the fraction of FIR colour selected AGN-candidates is $\backsim 28\%$ with a lower 
limit of $\backsim 7\%$ Our final catalogue contains 89 AGN 
candidates and is publicly available on the AMIGA web page 
(http://www.iaa.csic.es/AMIGA.html). A comparison with the results from the 
literature shows that the  AMIGA sample has the lowest ratio of AGN candidates, 
both globally and separated into early and late types. Field galaxies as well 
as poor cluster and group environments show intermediate values, while the 
highest rates of AGN candidates are found in the central parts of clusters and 
in pair/merger dominated samples. For all environments, early-type galaxies 
show a higher ratio of radio-excess galaxies than late types, as can be 
expected, since massive elliptical galaxies are the usual hosts of powerful 
radio continuum emission.}
   {We conclude that the environment plays a crucial and direct role in 
triggering radio nuclear activity and not only via the density-morphology 
relation. Isolated, early-type galaxies show a particularly low level of 
activity at radio wavelengths hence constituting the most nurture-free 
population of luminous early-type galaxies.}


   \maketitle
%

\section{Introduction}

Galaxy evolution depends strongly on the environment. In particular,
galaxy-galaxy interactions can induce nuclear activity by removing angular
momentum from the gas and, in this way, feeding the central black hole. Hence, a
higher rate of nuclear activity would be expected in interacting galaxies.
However, different studies of this topic lead to contradictory results. Some
works conclude that galaxies hosting an active galactic nucleus (AGN) have a
higher number of companions than those with nonactive galactic nuclei
\citep{Petrosian1982,Dahari1985,MacKenty1989,MacKenty1990,Rafanelli1995,Alonso2007},
 while  others do not find this excess of interacting companions or
find it only marginally \citep{Bushouse1986,Laurikainen1995,Schmitt2001a}.
\citet{Miller2003} find that the fraction of AGN is independent of the
environment even in clusters. \citet{Fuentes-Williams1988} found only a marginal
excess of similarly-sized galaxies but a clear excess of faint companions for
Seyfert galaxies. Most recent works find a different result depending on the
type of Seyfert galaxy \citep{Dultzin-Hacyan1999,Krongold2003}. Recently
\citet{Alonso2007} found for a sample of isolated galaxies a lower fraction of
Type 2 AGN (23\%) than for close galaxy pairs (30\%). The proportion of galaxies hosting an AGN in
extreme environments as compact groups was reported by \citet{Coziol2000} 
to be 70\%, while 50\% was reported by \citet{Shimada2000}. More recently
\citet{Geli2006a,Geli2006b} studied a sample of 215 compact groups
from the UZC (Updated Zwicky Catalog), and 42 galaxies belonging to compact groups from the Hickson Catalogue. They found in the first case  43\% of AGN and 20\% of transition
objects (objects with spectroscopic properties between pure AGN and pure star forming), 
while in the second subset  57\% showed characteristics consistent with
low-luminosity AGN with a low-ionization nuclear emission-line region (LINER) type 
and 16\% were transition objects.

Selection of AGN candidates using the radio-FIR correlation is also found in the
literature. This correlation is very tight and can be
used to distinguish galaxies for which their radio continuum emission is due
to star formation and which follow the correlation, and those with an AGN 
causing an enhancement of the  radio continuum emission that
lie above the correlation. 
It is known that all AGN are radio sources at some level, e.g.,
\citet{Ho2001} find that 85\% of the nuclei of Seyfert galaxies are detected at
radio wavelengths, with a wide range of intensities and morphologies (from
compact cores to jet-like features).
According to \citet{Reddy2004} a significant fraction of radio-excess objects
are associated with luminous AGN. For their sample of 114 galaxies in
nearby clusters, they find that 70\% of the radio-excess galaxies are AGN based on
different indicators, such as the presence of radio jets, X-ray emission, or optical
emission lines. \citeauthor{Reddy2004} consider this percentage as a lower limit 
of AGN among the radio-excess galaxies. They also analyse the
sample of \citet{Miller2001} and find that 80\% of the radio-excess galaxies in
their sample of local Abell clusters are spectroscopically-confirmed AGN. Far-infrared (FIR)
colours have also been demonstrated as useful to identify AGN candidates
\citep{deGrijp1985}. We discuss studies based on 
these methods in more detail 
in Sect.~\ref{sec:discussion} of this paper.

The contradictory results reached in previous studies might be due to the
design of the surveys, sometimes focused on galaxies with emission lines, or due to
different selection criteria of the samples. For example, although the sample in
\citet{Alonso2007} was selected with a well-defined isolation criterion (i.e., no
companions within a radius of 100 kpc and a velocity difference of 350 km/s) these
parameters are not restrictive enough 
to ensure that a galaxy has been isolated for a significant
fraction of its life. Hence, a well-defined sample of really isolated galaxies, which
have remained isolated for a significant part of their life, is needed.
The goal of the AMIGA project \citep[Analysis of the interstellar Medium of
Isolated GAlaxies, http://www.iaa.csic.es/AMIGA.html;][]{Verdes-Montenegro2005}
is, therefore, to identify a statistically-significant sample of the most isolated galaxies
in the local Universe and to quantify the properties of the interstellar medium
in these galaxies and its relationship to the star formation and nuclear
activity. In this paper, we will concentrate on the radio and FIR properties of this
sample.

We define our sample of isolated galaxies in Sect.~\ref{sec:sample}, and, in
Sect.~\ref{sec:data}, we describe the data. Then we make use of
different methods to select the AGN-candidates (Sect.~\ref{sec:methods}). In
Sect.~\ref{sec:catalogue}, we present the final catalogue of AGN-candidates, and
in Sect.~\ref{sec:discussion}, we discuss and compare our results to other studies
from the literature. Finally, we present our conclusions in
Sect.~\ref{sec:conclusions}.

\section{The sample}
\label{sec:sample}

The starting sample for the AMIGA project is based on the
Catalogue of Isolated Galaxies \citep[CIG;][]{Karachentseva1973} which is
composed of 1050 galaxies. In previous works we have:
1) revised all of the CIG positions \citep{Leon2003};
2) optically characterised the sample \citep{Verdes-Montenegro2005};
3) performed a revision of the morphologies \citep{Sulentic2006};
4) derived  mid-infrared (MIR) and FIR basic properties \citep{Lisenfeld2007};
5) performed a careful reevaluation of the degree of isolation of the CIG
\citep{Verley2007a,Verley2007b}; and
6) derived radio continuum properties \citep{Leon2007}.

In order to reduce biases in our statistical study we have used the completeness
test $<\mathrm{V}/\mathrm{Vm}>$, as explained in \citet{Verdes-Montenegro2005} and
\citet{Lisenfeld2007}.
We adopted $\mathrm{m_B}=15.0$ as the cutoff magnitude necessary to have a
reasonably complete sample. This subsample contains 719 galaxies and 
in this paper we will refer to it as 
 the ''complete subsample'', while the 1050 galaxies will be
referred to as the ''total sample''. For studies that use IRAS satellite data, 
the total sample is reduced to 1030 galaxies while the complete
subsample contains only 710 galaxies due to a number of galaxies falling in the 
region uncovered by IRAS known as the ''IRAS gap'' \citep[see][]{Lisenfeld2007}. 
This difference in the number of galaxies does
not affect the completeness of the subsample.

\section{The data}
\label{sec:data}

We performed our study using archive data we reprocessed as well as data
found in the literature.

\subsection{Observed and reprocessed data}

We obtained the FIR data for 1030 galaxies of the total sample by
reprocessing the data of the IRAS satellite 
with the SCANPI tool \citep{Lisenfeld2007}. We
obtained a better detection rate and an improved signal-to-noise level than in
previous IRAS catalogues.

The radio continuum data has been obtained from two different sources: a) NRAO
VLA Sky Survey (NVSS, 1.4 GHz; spatial resolution 45\arcsec) and b) Faint Images
of the Radio Sky at Twenty-cm (FIRST, 1.4 GHz; spatial resolution 5\arcsec) as
explained in detail in \citet{Leon2007}. The radio continuum fluxes used in this
paper were either taken from the NVSS survey catalogue or derived using
the original survey data and extracting the flux with SExtractor \citep{Bertin1996}
within a radius of 35\arcsec, obtaining in the latter case a higher detection
rate. We use NVSS fluxes because of the high detection rate and sensitivity of
this survey, which contains all galaxies belonging to our sample ($n=1050$). In
those cases where we found a radio enhancement (see Sect. \ref{sec:error}), we
complemented the NVSS data with the higher resolution images provided by
FIRST. In this way, we improved the spatial location of the
radio continuum emission to check whether this emission is due to a nuclear source
or to a projected, unrelated source.

We used the distances given in \citet{Verdes-Montenegro2005} calculated 
with the Hubble constant $H_{\mathrm{o}}=75\:\mathrm{km\:s^{-1}\:Mpc^{-1}}$.

We have computed the FIR and radio continuum luminosities with the following relations:
$$ \log L_{1.4 \mathrm{GHz}}(\mathrm{W~Hz^{-1}})=20.08 + 2 \log D + \log S_{1.4 \mathrm{GHz}}$$
$$ \log L_{60 \mathrm{\mu m}}(\mathrm{L_{\odot}})=6.014 + 2 \log D + \log S_{60 \mathrm{\mu m}} $$
\noindent where $D$ is the distance of the galaxy in Mpc and
$S_{1.4 \mathrm{GHz}}$ and $S_{60 \mathrm{\mu m}}$
are the flux densities in Jy ($1 \:\mathrm{Jy}= 10^{-26}\mathrm{W~m^{-2}~Hz^{-1}} $). 
The FIR luminosity \citep[$L_{\mathrm{FIR}}$;][]{Helou1988} 
is related to the $L_{60 \mathrm{\mu m}}$ by this formula:
$$ L_{\mathrm{FIR}}(\mathrm{L_{\odot}})=\left( 1+ \frac{S_{100 \mathrm{\mu m}}}{2.58 S_{60 \mathrm{\mu m}}} \right) L_{60 \mathrm{\mu m}}$$
The distribution of FIR luminosities for our sample peaks in
$\log(L_{\mathrm{FIR}}/\mathrm{L_{\odot}})$ = 9.5--9.75 \citep[see][]{Lisenfeld2007} and practically
all galaxies have FIR luminosities between $\log(L_{\mathrm{FIR}}/\mathrm{L_{\odot}}) = 7.5$
and $\log(L_{\mathrm{FIR}}/\mathrm{L_{\odot}}) = 11.25$. The bulk of the FIR luminosities
(98\%) lies below $\log(L_{\mathrm{FIR}}/\mathrm{L_{\odot}}) = 10.5$.

\subsection{Data from the literature}
\label{sec:lit}

We have cross-correlated our sample with existing databases of active galaxies,
in particular, the NASA Extragalactic Database (NED) \footnote{http://nedwww.ipac.caltech.edu/} and
the V\'eron-Cetty Catalogue of Quasars and Active Nuclei
\citep[12th Edition]{Veron-Cetty2006}.

The NED contains information on
the type of nuclear activity for $n=77$ galaxies of our sample. The
different classifications found in the sample are: LINER, HII,
Starburst, DANS (dwarf amorphous nuclear starburst), SBNG (small,
bright nucleus galaxy), NLAGN (narrow line AGN) and Seyfert. In some
cases, the information on the Seyfert type is also given. The
NLAGN are a mixture of Seyfert 2's, LINERs, and starburst/AGN
composites. HII and starburst are not taken into account in our analysis
since they are not considered non-stellar activity, but we list them in
Table~\ref{table:literature}. A total of $n=16$ galaxies from the AMIGA
sample are classified as Seyferts, 1 as AGN and 5 as NLAGN in their catalogue.

The V\'eron-Cetty Catalogue of Quasars and Active Nuclei is a very
complete compilation of active galaxies and quasars. It includes
position and redshift as well as photometry (U, B, V bands) and 6 and
21 cm flux densities when available. We have found $n=25$ of our
galaxies in this catalogue. A total of  $n=18$
are classified as Seyfert galaxies, 3 as HII, 3 as LINER and one has no
assigned class. The 3 HII galaxies are included in this catalogue
because in a previous version they were classified as
Seyfert galaxies.

The LINERs are also known as Sy3 (Seyfert 3) in these catalogues. Although
recent studies suggest some LINERs to be low-luminosity AGN
\citep{Ho1999,Omaira2006} this topic is not clear yet. In our study we consider
LINERs separately from other kinds of active galaxies.

In Table~\ref{table:literature} we list those galaxies found in the literature
to show nuclear emissions. The first column is the CIG number, the
second column is the type of emission found in NED, and the third column the one
in the V\'eron-Cetty catalogue. In the fourth column, we indicate the
classification taken for our study. It is obtained discarding those cases where a
disagreement is found between the NED and V\'eron-Cetty classification.

\begin{table}
\caption{Galaxies from AMIGA sample listed as active in the literature$^{1}$.}
\label{table:literature}
~\\
\centering
\begin{tabular}{c c c c c}
\hline \hline
CIG$^{1}$ & NED$^{2}$ & V-C$^{2}$ & Our study$^{2,3}$ \\
\hline
6       & HII           &       & HII \\
33      & Sy? LINER     & Sy?   & \\
45      & HII           &       & HII \\
55      & LINER HII     &       & \\
56      & Sbrst         &       & Sbrst \\
57      & Sy2           & Sy2   & Sy2 \\
....     & ...           & ...   & ... \\
\hline
\end{tabular}
\begin{list}{}{}
\item[$^{\rm 1}$] Columns: 
1) CIG number; 
2) NED classification; 
3) V\'eron-Cetty Catalogue classification;
4) Classification adopted for our study.
\item[$^{\rm 2}$]
Sy = Seyfert;
HII = nuclear HII region;
Sbrst= Starburst;
SBNG = small, bright nucleus galaxy;
NLAGN = narrow line active galaxies (a mixture of Seyfert 2's, LINERs, and starburst/AGN composites);
DANS = dwarf amorphous nuclei starburst galaxies
\item[$^{\rm 3}$] Classification obtained discarding
those cases where a disagreement was found between the NED and V\'eron-Cetty classification.
\end{list}
\end{table}

\section{AGN selection methods}
\label{sec:methods}

Several methods to select AGN candidate galaxies exist. These
different selection methods are generally biased toward a certain
kind of nuclear activity. In this section, we discuss the results of
applying selection criteria based on combined radio continuum and FIR data.

\subsection{Radio-FIR correlation}
\label{sec:radio-FIR}

The correlation between the FIR and the radio continuum emission
is very tight and is attributed to star formation
\citep{Helou1985,Condon1991}. Massive stars ($ M \gtrsim 8 \mathrm{M_{\odot}}$)
heat the dust which re-emits in the FIR. On the other hand, the supernovae produced
at the end of the lives of these short-lived stars are ultimately responsible for
the radio synchrotron emission from cosmic ray electrons accelerated in their
shocks. Deviation from this correlation may be produced by nuclear activity
\citep{Sopp1991,Niklas1997,Yun2001,Drake2003}, since those AGN with compact (200
pc) radio cores, i.e., some radio-quiet AGN and all radio-loud AGN
\citep[e.g.,][]{Impey1993,Roy1998} exhibit a radio excess.

Radio-loud active galaxies amount to approximately 10\% of the 
optically-identified galaxies with an AGN \citep{Kellermann1989,Hooper1995}. In radio-loud
AGN, the radio emission from the nucleus is added to the emission produced by
star formation. If this additional emission is strong enough, an excess of radio
emission will be found with respect to the radio-FIR correlation. We use the
radio-FIR correlation to identify radio AGN candidates.

Since 468 galaxies in our complete subsample have at least one upper limit
(radio or FIR) we have used survival analysis to compute the correlation to
exploit the information carried in the upper limits. The Schmitt method
\citep{Schmitt1985,Isobe1986} allows us to compute a correlation when upper limits
exist in both the dependent and the independent variables. We use the
complete subsample to calculate the correlation. We computed two regression
lines to compare with other regressions of different samples and
authors, one for $\log L_{1.4 \mathrm{GHz}}(\mathrm{W~Hz^{-1}})$ versus $\log
L_{60 \mathrm{\mu m}}(\mathrm{L_{\odot}})$ and the other for $\log L_{1.4
\mathrm{GHz}}(\mathrm{W~Hz^{-1}})$ versus $\log L_{\mathrm{FIR}}(\mathrm{L_{\odot}})$.
The regression lines are:
$$ \log L_{1.4 \mathrm{GHz}}(\mathrm{W~Hz^{-1}})= [1.02 \pm 0.03] \log (L_{\mathrm{FIR}}/\mathrm{L_{\odot}}) + [11.4 \pm 0.3]$$
$$ \log L_{1.4 \mathrm{GHz}}(\mathrm{W~Hz^{-1}})= [1.025 \pm 0.023] \log (L_{60 \mathrm{\mu m}}/\mathrm{L_{\odot}}) + [11.75 \pm 0.21].$$

In both cases, the slope is close to one. These regression lines have been
computed adopting $\log L_{1.4 \mathrm{GHz}}$ as the independent variable and 
minimising the residuals in the y-axis. This is the usual way of computing the
radio-FIR correlation in the literature, which enables us to
carry out comparisons.
 Since we are interested in the
physical relation between the two variables we prefer to use a symmetric method,
the bisector best-fit \citep{Isobe1990}. In this case the 
regression lines are:
$$ \log L_{1.4 \mathrm{GHz}}(\mathrm{W~Hz^{-1}})= [1.06 \pm 0.03] \log (L_{\mathrm{FIR}}/\mathrm{L_{\odot}}) + [11.1 \pm 0.3]$$
$$ \log L_{1.4 \mathrm{GHz}}(\mathrm{W~Hz^{-1}})= [1.07 \pm 0.03] \log (L_{60 \mathrm{\mu m}}/\mathrm{L_{\odot}}) + [11.31 \pm 0.23],$$
\noindent hence deviating slightly  from linearity.

\subsubsection{Radio-excess galaxies}
\label{sec:radio-excess}

\begin{figure}
\centering
\resizebox{9.5cm}{9.5cm}{\includegraphics{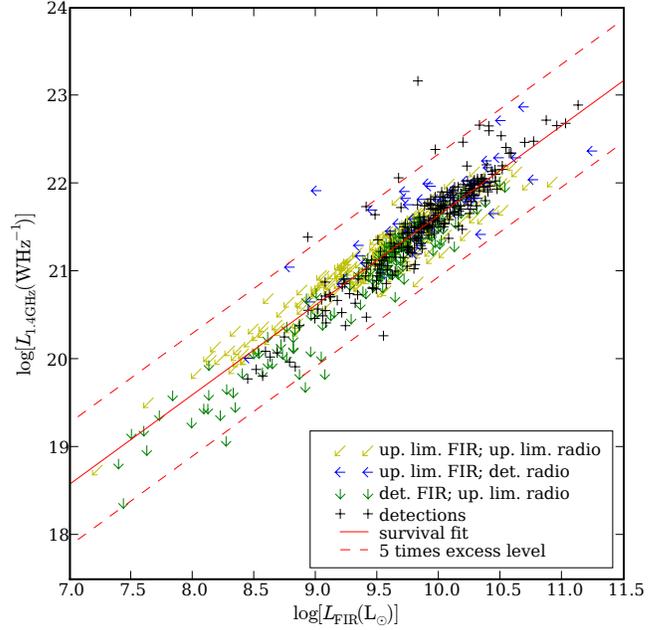}}
   \caption{Radio versus FIR luminosity for the complete subsample \mbox{($n=710$)}. We show the correlation
as a solid line and the 5 times radio-excess and FIR-excess levels as dashed lines. The galaxies above the 
upper-dashed line are the radio-excess galaxies.
           }
      \label{fig:l}
\end{figure}

Radio-excess galaxies are defined, according to
\citet{Yun2001},  as those whose radio luminosity is larger
than 5 times the value predicted by the radio-FIR correlation. 
In Fig.~\ref{fig:l}, we show the galaxies of our complete
subsample with the regression fit plotted as a solid line. The dashed lines
denote a deviation by a factor 5. Galaxies above the upper-dashed line
are selected as radio-excess galaxies. There are 6 radio-excess galaxies in the
complete subsample ($n=710$) and 2 more in the total sample. Other
studies \citep{Niklas1995,Condon2002,Mauch2007} use a cutoff factor of 3 
for the radio excess, hence this cutoff will be also used 
to allow comparison with these samples.
Using this value for our sample, we find 16 radio-excess galaxies in the
complete subsample and 4 more in the total sample. In
Table~\ref{tabla:exceso-l}, we list the galaxies with a radio-excess (of a 
factor 3) for the
total sample and, in Table~\ref{tabla:exceso}, we detail separately the number of
radio-excess galaxies and percentages for the complete subsample and for
the early and late-type galaxies. The percentages are normalised to the 
number of galaxies that can be classified using this method. We can classify galaxies with 
detections in both bands, galaxies detected in radio continuum with an upper limit in FIR 
above the selection line and galaxies detected in FIR with an upper limit in radio continuum
below the selection line.

None of the galaxies classified as Seyfert, LINER, NLAGN or AGN as listed in column~4 of
Table~\ref{table:literature} show a radio-excess with respect to the
radio-FIR correlation by a factor 5 or more. 
We find 3 Seyfert galaxies located by more than a factor of 3 above the
FIR-radio correlation:
\mbox{CIG 72}, 692 and 877. \mbox{CIG 72} has an upper limit in FIR and,
according to \mbox{J. Lim} (priv. comm.), shows a tidal tail in HI linking it to a small
companion. \mbox{CIG 692} has been observed with the VLA in its A configuration
at 8.3~GHz by \citet{Schmitt2001b} finding a symmetric triple source with a total
extent of 3 kpc. \mbox{CIG 877} is an elliptical galaxy studied by
\citet{Marcum2004}. They suggest that the IRAS emission assigned to this galaxy
might be produced by two nearby stars. If this is true the radio excess would be
even higher.

\begin{table}
\caption{Radio-excess galaxies found using the radio-FIR correlation.}
\begin{center}
\label{tabla:exceso-l}
\begin{tabular}{c c c c c c}
\hline \hline
CIG & $\log L_{\mathrm{FIR}}$ $^{1}$ & Code$^{2}$ &  $\log L_{1.4 \mathrm{GHz}}$ $^{3}$ & T$^{4}$ & Excess factor$^{5}$ \\
\hline
\multicolumn{6}{c}{Galaxies belonging to the complete subsample.}\\
\hline
41    &  9.67   &  0  &  22.07  & Sc    & 5.7 \\
72    &  9.72   &  1  &  21.91  & Sbc   & 3.6 \\
97    &  10.19  &  0  &  22.47  & Sc    & 4.3 \\
156   &  9.41   &  0  &  21.74  & SBab  & 5.0 \\
187   &  9.83   &  0  &  23.17  & Sc    & 50.2\\
248   &  10.40  &  0  &  22.60  & Sc:   & 3.5 \\
287   &  9.97   &  0  &  22.39  & SBbc  & 6.0 \\
480   &  9.48   &  0  &  21.65  & S0a   & 3.4 \\
488   &  10.50  &  1  &  22.72  & Sb    & 3.7 \\
571   &  8.93   &  0  &  21.39  & Sc    & 6.9 \\
591   &  9.46   &  1  &  21.39  & Sbc   & 4.1 \\
692   &  10.33  &  0  &  22.67  & Sb    & 4.9 \\
734   &  8.79   &  1  &  21.05  & S0a   & 4.4 \\
877   &  10.40  &  0  &  22.66  & E:    & 4.0 \\
893   &  10.69  &  1  &  22.87  & E/S0  & 3.4 \\
1045  &  8.78   &  1  &  21.92  & S0    & 19.7\\
\hline
\multicolumn{6}{c}{Galaxies not belonging to the complete subsample.}\\
\hline
57    &  10.23  &  1  &  22.73  & Sb    & 5.6 \\
510   &  9.42   &  1  &  21.61  & Sc    & 3.6 \\
836   &  10.17  &  1  &  23.29  & E/S0  & 29.7\\
999   &  9.90   &  0  &  22.09  & Sa    & 3.5 \\
\hline
\end{tabular}
\begin{list}{}{}
\item[$^{\rm 1}$] $\log L_{\mathrm{FIR}}$ from \citet{Lisenfeld2007}.
\item[$^{\rm 2}$] Detection code for $\log L_{\mathrm{FIR}}$: 0 for detections and 1 for upper limits.
\item[$^{\rm 3}$] $\log L_{1.4 \mathrm{GHz}}$ from \citet{Leon2007}.
\item[$^{\rm 4}$] Morphological type from \citet{Sulentic2006}, except for \mbox{CIG 999} \citep[NED]{Karachentseva1973}.
\item[$^{\rm 5}$] Radio luminosity excess above the radio-FIR correlation.
\end{list}
\end{center}
\end{table}

\begin{table}
\caption{Radio-excess ratios for the complete subsample.$^1$}
\begin{center}
\label{tabla:exceso}
\begin{tabular}{|c|c |c |c |c|c| c |c| c|}
\hline \hline
\multirow{3}{*}{Morpho$^3$} & \multicolumn{4}{|c|}{Radio-excess galaxies} & \multicolumn{4}{|c|}{FIRST revised $^2$} \\
\cline{2-9}
 & \multicolumn{2}{|c|}{Factor 5} & \multicolumn{2}{|c|}{Factor 3} & \multicolumn{2}{|c|}{Factor 5} & \multicolumn{2}{|c|}{Factor 3} \\
\cline{2-9}
 & N & \% & N & \% & N & \% & N & \% \\
\hline
All T ($n=397$) & 6 & 1.5 & 16 & 4.0 & 3 & 0.8 & 12 & 3.0 \\
E-S0a ($n=21$)  & 1 & 4.8 & 4  & 19.1 & 1 & 4.8 & 4  & 19.1 \\
Sa-Irr ($n=376$) & 5 & 1.3 & 12 & 3.2 & 2 & 0.5 & 8  & 2.1 \\
\hline
\end{tabular}
\begin{list}{}{}
\item[$^{\rm 1}$] All percentages for the fraction of radio-excess galaxies are upper limits as explained in Sect.~\ref{sec:error}.
\item[$^{\rm 2}$] In Sect.~\ref{sec:error} we explain how we revise these numbers using the FIRST
survey.
\item[$^{\rm 3}$] Morphological subsamples. The percentages are computed over the number of galaxies for each subsample.
\end{list}
\end{center}
\end{table}

\subsubsection{q-parameter}
\label{sec:q}

The q-parameter \citep{Helou1985} is a good estimator of the deviation from the
radio-FIR correlation. It has been found to be independent of the starburst
strength \citep{Lisenfeld1996} and distance \citep{Yun2001}. It is defined as:
$$q \equiv \log \left[ \frac{\mathrm{FIR}}{3.75 \cdot 10^{12} \mathrm{W m^{-2}}} \right] - \log\left[ \frac{S_{1.4 \mathrm{GHz}}}{\mathrm{W m^{-2} Hz^{-1}}} \right]$$
where $ S_{1.4 \mathrm{GHz}} $ is the flux density at 1.4 GHz in units of 
$\mathrm{W m^{-2} Hz^{-1}}$ and FIR is the FIR flux calculated as:
$$ \mathrm{FIR} \equiv 1.26 \cdot 10^{-14} \left( 2.58 S_{60\mathrm{\mu m}} + 
S_{100\mathrm{\mu m}}\right)
\mathrm{W m^{-2}}$$
where $S_{60\mathrm{\mu m}}$ and $ S_{100\mathrm{\mu m}}$ are IRAS 
$60\mathrm{\mu m}$ and $100\mathrm{\mu m}$ band flux densities in Jy.

\begin{figure}
\centering
   \resizebox{9.5cm}{9.5cm}{\includegraphics{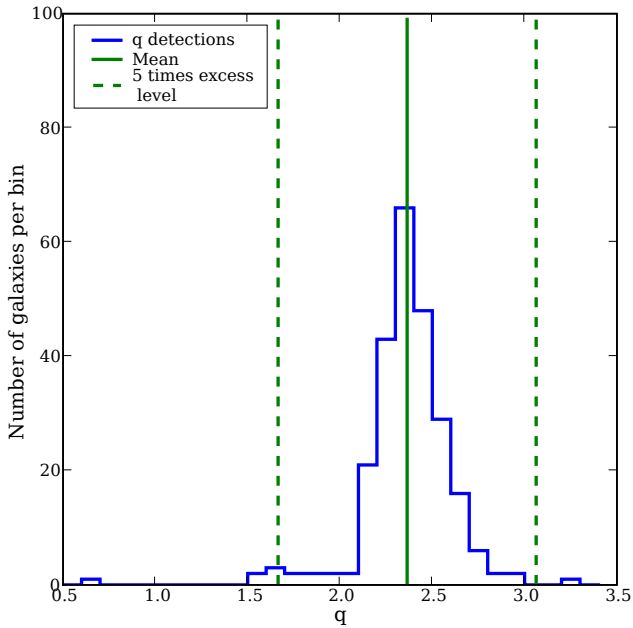}}
   \caption{Histogram of the q-parameter showing the galaxies of the complete subsample detected both in radio and infrared ($n=248$). We also show the mean with a solid line and the 5 times radio-excess and FIR-excess limits with a dashed line. Radio-excess galaxies are located in the left tail of the histogram with lower q values.
           }
      \label{fig:q}
\end{figure}

We have computed the q-parameter using only the galaxies detected both in FIR
and in radio continuum ($n = 248$) because non-detections will result in upper 
and lower limits of the q-parameter and, as far
as we know, there is no statistical method to take both of them into account 
simultaneously. In Fig.~\ref{fig:q}, a histogram of the q-parameter is shown. The
mean value is 2.36 with a dispersion of $\sigma = 0.24$. We define radio-excess
galaxies in the same way as before by a deviation of more than a factor 5, which
translates into $q < 1.66$. This condition is fullfilled for 4 out of 248 galaxies
(\mbox{CIG 41}, 187, 287 and 571), corresponding to 1.6\% of the studied galaxies. 
This rate is slightly higher than the one
derived from the luminosity correlation ($\approx 1\%$), but we have to take
into account that the errors are very high (small numbers) and that radio-excess
galaxies probably have a higher chance of being detected in both radio continuum
and FIR. So this fraction could be considered as an upper limit for the fraction
of radio-excess galaxies.

\subsubsection{Projected radio sources}
\label{sec:error}

Since the spatial resolution of the NVSS is not very high with respect to the
size of the galaxies in our sample \citep[97\% of them are smaller than
4\arcmin, cf. ][]{Lisenfeld2007} there is a chance that the radio excess found
using the NVSS data is in fact due to a background/foreground source projected
in the line of sight of the galaxy. We computed this probability with two
methods. In the first method, 
we estimated the density of radio sources expected in the NVSS dividing
the number of sources in the catalogue by the covered area, and obtained a value
of 159825.4 sources/sr. Since we have searched for a NVSS source within a radius
of 35\arcsec\ of each CIG galaxy, we determine that we could expect $n \approx 15$
unrelated sources in our total sample. In order to refine the previous value
with a more local estimation of the probability, we estimated the average
density of NVSS radio sources within a radius of 5\degr\ of each CIG galaxy. The
mean value found for the density is similar to the previous one within 10\%
(172561.7 sources/sr), hence implying $n \approx  16$ unrelated sources in our
complete sample. The radio flux of a source projected in the field of a CIG
galaxy would add to its normal emission and take it above the radio-FIR
correlation. Since the expected number of unrelated sources is of the same order
as the number of radio-excess galaxies, we expect an even lower ratio of
galaxies above the radio-FIR correlation.

In the second method, we have made use of the higher spatial resolution 
of the FIRST survey to check whether the radio emission of the radio-excess 
galaxies is associated with the nucleus of the galaxy. Three among
the 6 galaxies in the complete subsample with a radio excess larger than a
factor 5 are covered by FIRST and all of them turn out to be sources unrelated to
the nuclei. Seven of the 16 galaxies with a radio excess larger than a factor 3
in the same sample are covered by FIRST, and 4 of them are also unrelated sources. One of
the 4 galaxies showing a radio excess not belonging to the complete subsample is
covered by FIRST and we find its radio emission  also to be separated from the nucleus. In
Table~\ref{table:FIRST}, we list for all these sources the NVSS and FIRST fluxes
and the distance between the nucleus  and the closest FIRST
source. NVSS and FIRST data were taken on different dates, but the fluxes
in both surveys are very similar so that we can discount the possibility that the extranuclear emission
is produced by a supernova. With these results, we have revised the percentages
of radio-excess galaxies and give them in Table~\ref{tabla:exceso}.

Since 5 out of the 8 galaxies revised with FIRST have proven to be projected 
sources and given the number of unrelated sources that we have 
estimated, there is a high probability that 
many of the remaining radio-excess galaxies are also unrelated 
sources. We computed the revised fractions of radio-excess galaxies 
considering all the galaxies without FIRST data as genuine detections, 
consequently the derived fractions (Table~\ref{tabla:exceso}) should be considered as 
upper limits.

\begin{table}
\caption{Radio-excess galaxies in FIRST.}
\label{table:FIRST}
~\\
\centering
\begin{tabular}{c c c c}
\hline \hline
CIG &  NVSS flux (mJy) & FIRST flux$^{1}$  (mJy) & Distance$^{2}$ (\arcsec) \\
\hline
  187  & 97.9 & 90.30 & 18.3 \\
  248  & 14.3 & 10.24 &  0.5 \\
  287  & 15.7 & 11.42 & 30.3 \\
  480  &  6.1 &  5.56 &  0.3 \\
  510  &  4.6 &  2.61 & 16.5 \\
  571  & 16.3 & 13.88 & 36.9 \\
  591  &  5.1 &  6.54 &  0.6 \\
  734  &  3.3 &  2.11 & 26.8 \\
\hline
\end{tabular}
\begin{list}{}{}
\item[$^{\rm 1}$] Sum of FIRST fluxes within the NVSS search radius if more than one source is present.
\item[$^{\rm 2}$] Distance from the CIG galaxy centre to the FIRST source. Distance to the nearest FIRST source if more than one source is present.
\end{list}
\end{table}

\subsection{IRAS colour}
\label{sec:IRAS-colours}

In the work of \citet{deGrijp1985} a method to identify AGN candidates
using FIR properties was presented. Galaxies hosting an AGN have, in
general, a flatter spectrum in FIR. This is due to the warmer
temperatures of the dust heated by the central engine. The advantage
of the method is that it can identify obscured AGN that cannot be
found with other wavelengths or methods. The success rate of this
method is about 70\%. We select those galaxies 
with $S_{25 \mathrm{\mu m}}/S_{60 \mathrm{\mu m}} \geq 0.18$ as 
AGN-candidates, following the studies by \citet{Yun2001} and 
\citet{Reddy2004}.

\begin{figure}
\centering
   \resizebox{9.5cm}{9.5cm}{\includegraphics{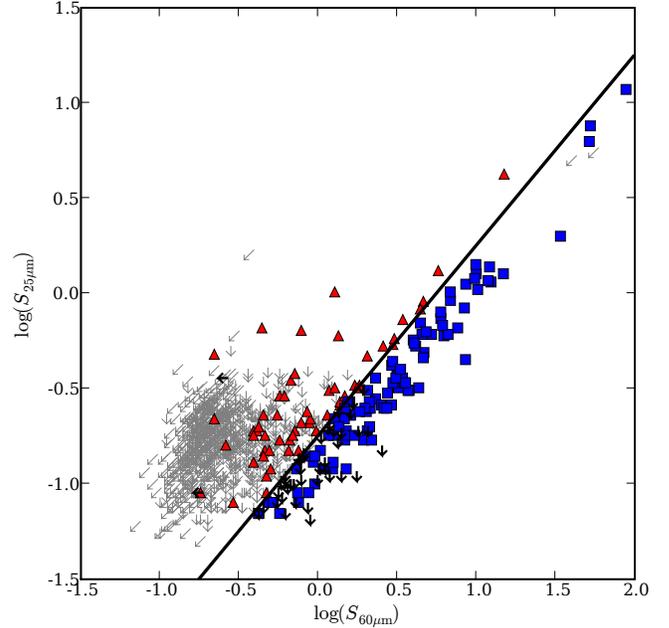}}
   \caption{Plot of $\log S_{25 \mathrm{\mu m}}$ versus $\log S_{60 \mathrm{\mu m}}$ for the 
total sample ($n=1030$). The solid line corresponds to 
$S_{25 \mathrm{\mu m}}/S_{60 \mathrm{\mu m}} = 0.18$. 
Galaxies classified as AGN-candidates lie above this line and are plotted as triangles 
and black left-arrows. Galaxies classified as non AGN-candidates are below the line, 
plotted with squares and black down-arrows. The remaining galaxies can not be classified 
due to upper limit in the fluxes (grey arrows).
           }
      \label{fig:iras_selection}
\end{figure}

In Fig.~\ref{fig:iras_selection}, we plot $\log S_{25 \mathrm{\mu m}}$ versus
$\log S_{60 \mathrm{\mu m}}$ for the total sample. The $S_{25 \mathrm{\mu
m}}/S_{60\mathrm{\mu m}}$ ratio of 0.18 is plotted as a solid line. The 
triangles and squares
denote detections and the arrows upper limits. A total of 197 galaxies of the
total sample or 162 of the complete subsample can be classified using this
method (Table~\ref{tabla:IRAS}) since they are: a) detections in both bands; b)
detections at $S_{60 \mathrm{\mu m}}$ and upper limits at $S_{25 \mathrm{\mu
m}}$ with flux ratios below the $S_{25 \mathrm{\mu
m}}/S_{60 \mathrm{\mu m}} = 0.18$ line; or 
c) detections at $S_{25 \mathrm{\mu m}}$ and upper limits at $S_{60
\mathrm{\mu m}}$
with flux ratios above the $S_{25
\mathrm{\mu m}}/S_{60 \mathrm{\mu m}} = 0.18$ line. 
Fifty-eight galaxies in the total sample and 46 in the complete
subsample are AGN-candidates. Hence 28.4\% of the galaxies for which a
classification could be assigned are AGN-candidates. We obtained 
a lower limit to the total fraction of AGN-candidates of 6.5\% 
by normalising to the complete subsample ($n=710$). This is a lower limit
because it assumes that the unclassified galaxies do not host an AGN.

\begin{table}
\caption{Classified galaxies using the IRAS colour method.}
\label{tabla:IRAS}
~\\
\centering
\begin{tabular}{c c c c c c}
\hline \hline
CIG &  S25 (Jy)$^{1}$  & Code$^{1,2}$ & S60 (Jy)$^{1}$ & Code$^{1,2}$ & Classification$^{3}$ \\
\hline
\multicolumn{6}{c}{Galaxies belonging to the complete subsample.}\\
\hline
  4  &  0.61  &  0   & 5.19   & 0   & normal \\
 41  &  0.19  &  0   & 0.39   & 0   & AGN \\
 55  &  0.36  &  0   & 2.30   & 0   & normal \\
 56  &  0.23  &  1   & 1.60   & 0   & normal \\
 66  &  0.21  &  0   & 1.23   & 0   & normal \\
 ...  &  ...  &  ...   & ...   & ...   & ... \\
\hline
\multicolumn{6}{c}{Galaxies not belonging to the complete subsample.}\\
\hline
  26  &  0.48  &  0  &  0.22  &  0  & AGN \\
  62  &  0.25  &  0  &  2.07  &  0  & normal \\
  80  &  0.92  &  0  &  6.84  &  0  & normal \\
 105  &  0.66  &  0  &  7.60  &  0  & normal \\
 121  &  0.60  &  0  &  6.22  &  0  & normal \\
 ...  &  ...  &  ...   & ...   & ...   & ... \\
\hline
\end{tabular}
\begin{list}{}{}
\item[$^{\rm 1}$] Data from \citet{Lisenfeld2007}.
\item[$^{\rm 2}$] Detection code: 0 for detections and 1 for upper limits.
\item[$^{\rm 3}$] Normal for $S_{25 \mathrm{\mu m}}/S_{60 \mathrm{\mu m}} < 0.18$ 
and AGN for $S_{25 \mathrm{\mu m}}/S_{60 \mathrm{\mu m}} \geq 0.18$.
\end{list}
\end{table}

Four of the radio-excess candidate galaxies can be also classified with the IRAS colour method and 
all of them (CIG 41, CIG 248, CIG 692 and CIG 877) are found to be AGN candidates. These galaxies
have very warm IRAS colours with $S_{25 \mathrm{\mu m}}/S_{60 \mathrm{\mu m}}$ being higher than 0.45 
in all cases.

\begin{figure}
\centering
   \resizebox{9.5cm}{9.5cm}{\includegraphics{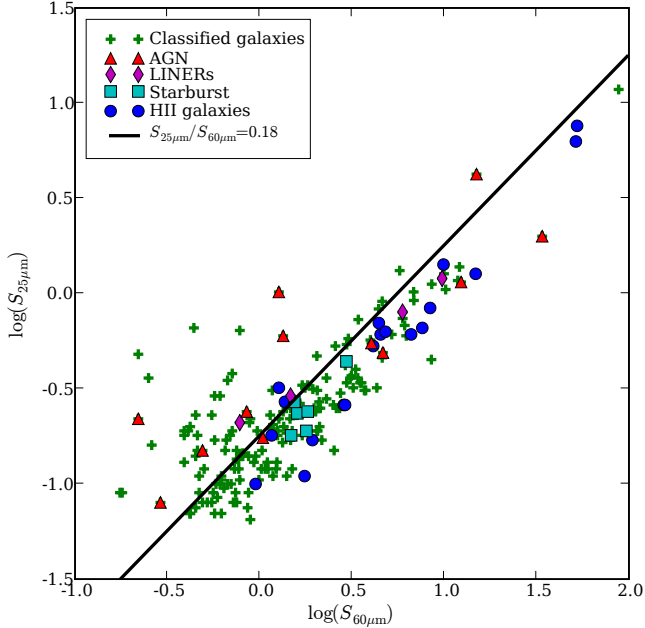}}
   \caption{Plot of $\log S_{25 \mathrm{\mu m}}$ versus $\log S_{60 \mathrm{\mu m}}$ 
for the CIG sample with available data about nuclear emission from the literature. 
The solid line corresponds to $S_{25 \mathrm{\mu m}}/S_{60 \mathrm{\mu m}} = 0.18$. 
Galaxies that can be classified using this 
method are plotted as crosses \mbox{($n=197$)}. AGN galaxies from the literature (triangles) 
are usually above the line. LINERs are plotted as diamonds. Most of the galaxies classified
as HII and starburst in the literature (circles and squares respectively) are below the selection line.
           }
      \label{fig:iras_lit}
\end{figure}

Since for a number of CIG galaxies a detailed classification of their nuclear
emission has been found in the literature (Sect.~\ref{sec:lit}), we investigate
specifically their location in the IRAS colour plot. This has allowed us to test
the accuracy of the classification method. In Fig.~\ref{fig:iras_lit}, we plot
the 197 galaxies that we were able to classify based on their IRAS colour and
flag those galaxies classified as AGN, HII, LINER or starburst in the
literature. A large fraction of the AGN galaxies are located above the selection cut
\mbox{($7/12\approx 60\%$)}, and almost all the HII and starburst galaxies
($23/25$) are below the line.  
This confirms our expectation that dust in AGN is usually warmer,
hence, showing a higher $S_{25 \mathrm{\mu m}}$ to $S_{60\mathrm{\mu m}}$ ratio
while galaxies classified as HII should have lower values for that ratio.
This result has no statistical significance, but shows the good
agreement between the classification of their nuclear emission and their
location in the $S_{25 \mathrm{\mu m}}$ versus $S_{60 \mathrm{\mu m}}$ plot.

\section{AGN catalogue}
\label{sec:catalogue}

We have compiled the results obtained from the methods explained in
Sect.~\ref{sec:methods} as well as the data from the literature to
produce a single catalogue of AGN-candidate galaxies for the total sample
(Table~\ref{table:catalogue}). In the catalogue, we also indicate the galaxies
composing the complete subsample. Since our confindence level varies as to the 
presence of an AGN depending on the selection method and its properties, 
as explained in the previous sections, we have included
information on the method used.

The final catalogue is composed of:
\begin{itemize}
\item \textit{Active galaxies found in the NED and the V\'eron-Cetty
catalogue (see Sect.~\ref{sec:lit}).} We have selected the 29 galaxies
classified as Seyfert, LINER, NLAGN, or AGN
as listed in column 4 of Table~\ref{table:literature}.
\item \textit{Radio-excess candidates galaxies selected with the
radio-FIR correlation (see Sect.~\ref{sec:radio-excess} and Sect.~\ref{sec:error}).} We have included the 15
galaxies with at least a factor 3 radio excess using the $\log L_{1.4 \mathrm{GHz}}(\mathrm{W~Hz^{-1}})$
versus $\log L_{\mathrm{FIR}}(\mathrm{L_{\odot}})$ correlation, taking into account the FIRST revision (Table~\ref{tabla:exceso-l}).
\item \textit{AGN-candidates selected with the IRAS colour criterion (see Sect.~\ref{sec:IRAS-colours}).}
We have added the 58 AGN-candidates selected using this method to the catalogue (Table~\ref{tabla:IRAS}).
\end{itemize}

\begin{table*}
\caption{Catalogue of AGN-candidates for the total sample$^{1}$.}
\label{table:catalogue}
~\\
\centering
\begin{tabular}{c c c c c c c c}
\hline \hline
\multirow{2}{*}{CIG} & \multirow{2}{*}{$\alpha$ (J2000)} & \multirow{2}{*}{$\delta$ (J2000)} & \multirow{2}{*}{Literature} & \multirow{2}{*}{FIR colour} & Radio-excess 3 & Radio-excess & Complete  \\
  & & & & & Factor 3  & Factor 5 & subsample \\
\hline
  26 &  00:31:52.85  &  +37:40:43.2  & - & AGN & 0 & 0 & 0 \\
  41 &  00:58:23.36  &  +36:43:50.2  & - & - & 1 & 1 & 1 \\
  44 &  01:06:35.60  &  +10:31:18.1  & - & AGN & 0 & 0 & 1 \\
  57 &  01:37:48.25  &  +02:17:27.3  & Sy2 & - & 1 & 1 & 0 \\
  69 &  01:53:42.23  &  +29:56:01.5  & - & AGN & 0 & 0 & 1 \\
  ... & ... & ... & ... & ... & ... & ... & ... \\
\hline
\end{tabular}
\begin{list}{}{}
\item[$^{\rm 1}$]
Columns: \\
(1) CIG number.\\
(2) $\alpha$ (J2000).\\
(3) $\delta$ (J2000).\\
(4) Classification in the literature as listed in column 4 of Table~\ref{table:literature}.\\
(5) AGN candidates using the FIR colour criterion.\\
(6) 1 if is a radio-excess candidate galaxy using a factor 3 cutoff, 0 if not.\\
(7) 1 if is a radio-excess candidate galaxy using a factor 5 cutoff, 0 if not.\\
(8) 1 if the galaxy belongs to the complete subsample, 0 if not.\\
\end{list}
\end{table*}

\section{Discussion}
\label{sec:discussion}

Based on the radio-FIR correlation we have determined a very low rate of AGN
candidates, which further decreases after rejection of projected sources with FIRST data
(Sect.~\ref{sec:error}). For the complete subsample, we find at most 0.8\% of
radio-excess galaxies with a factor 5 cutoff (or less than 4.0\% for a factor 3 cutoff;
Table~\ref{tabla:exceso}). The AGN candidate rate based on the FIR colour
selection is $\backsim 28\%$ (with a lower limit of $\backsim 7\%$), higher than the
radio-excess ratio  as expected, since the colour method is sensitive to all
types of AGN and not only to radio-loud objects. The optical luminosities of the
galaxies with a radio excess have the same mean value as the complete sample, and
the same occurs for FIR luminosities, although the results are only illustrative
due to their low number. A trend is found for radio excess galaxies to have
earlier morphological types, particularly in the subset showing a radio-excess
above the factor 3 cutoff: 25\% of the galaxies with radio-excess have
morphologies earlier than Sa, while for the complete sample the percentage of
early types is 12\%.

Our sample is mostly radio quiet, with only a few galaxies (1.3~\%) above a 
radio power of $10^{23}~\mbox{W~Hz}^{-1}$, consistent with the high ratio of 
late-type galaxies, which are usually radio quiet, in our sample. The only 
galaxy with a radio power above $10^{23}~\mbox{W~Hz}^{-1}$ 
in the complete subsample was found, based on
FIRST data, to be  an unrelated source (\mbox{CIG 187}). 
The radio continuum emission is dominated
by mild star-formation \citep{Leon2007}, in contrast to a sample dominated by
radio-AGN, the 2dF Galaxy Redshift Survey \citep[2dFGRS;][]{Sadler2002}, where
60\% of the galaxies are classified as AGN, and 75\%  have a radio power at 1.4
GHz higher than $10^{23}~\mbox{W~Hz}^{-1}$.  The CIG presents at most 30\% of
the total radio power density  at 1.4 GHz  of the  star-forming sample of
\citet{Yun2001} or 10\% of  AGN-dominated samples  \citep{Sadler2002,Condon2002}
in the local Universe. This low level of radio continuum emission at 1.4 GHz for
the CIG is confirmed by the low radio-to-optical ratio compared to 
emission-line galaxies  \citep[KISS sample;][]{VanDuyne2004}  as shown in
\citet{Leon2007}.

\subsection{Comparison with other samples}

Studies of the radio power  and FIR emission of galaxies have been performed
mostly for two kinds of samples: those referred to as field galaxies in the
literature
\citep[e.g.,][]{Condon1991b,Yun2001,Miller2001,Corbett2002,Condon2002,Drake2003}
where usually no environmental selection criterion has been applied, and cluster
samples  \citep[e.g.,][]{Niklas1995,Andersen1995,Miller2001,Reddy2004,Omar2005}.
We summarise the main results of these papers in Table~\ref{tabla:exceso-rate}
and~\ref{tabla:color-rate}.

\begin{table*}
\caption{Rate of radio-excess galaxies in the literature.}
\begin{center}
\label{tabla:exceso-rate}
\begin{tabular}{|l|l|c|c|c|c|c|c|c|c|c|c|}
\hline \hline
\multicolumn{1}{|c|}{} & \multicolumn{1}{|c|}{} & \multicolumn{9}{|c|}{Rate of Radio-excess galaxies$^{\rm 1}$} & \\
\cline{3-11}
\multicolumn{1}{|c|}{} & \multicolumn{1}{|c|}{} & \multicolumn{3}{|c|}{Total} & \multicolumn{3}{|c|}{E-S0a} & 
\multicolumn{3}{|c|}{Sa-Irr} & \\
\cline{3-11}
Sample  &  Environment  & N$^{\rm 2}$ & F5$^{\rm 3}$ & F3$^{\rm 3}$ & N$^{\rm 2}$ & F5$^{\rm 3}$ & F3$^{\rm 3}$ & N$^{\rm 2}$ & F5$^{\rm 3}$ & F3$^{\rm 3}$ & Notes \\
\hline
AMIGA$^{\rm 4}$ & Isolated & 397 & 0.8 & 4.0 & 21 & 4.7 & 19.1 & 376 & 0.5 & 2.1 &  $m_{B}<15$ \\
\multirow{2}{*}{Condon 1991}& \multirow{2}{*}{~~~~-} & \multirow{2}{*}{122}& \multirow{2}{*}{32.0} & \multirow{2}{*}{33.6} & \multirow{2}{*}{11} & \multirow{2}{*}{90.9} & \multirow{2}{*}{90.9} & \multirow{2}{*}{71} & \multirow{2}{*}{9.9} & \multirow{2}{*}{12.7} & $S_{60\mathrm{\mu m}}>
0.2 \mathrm{Jy}$ \\
 &  & &  &  & &  &  & &  &  &  and $S_{4.85\mathrm{GHz}}> 25 \mathrm{mJy}$ \\
Yun 2001 & ~~~~- & 1809 & 1.3 & - & - & - & - & - & - & - & $S_{60\mathrm{\mu m}}> 2 \mathrm{Jy}$\\
Corbett 2002 & ~~~~- & 82 & 2.4 & 4.9 & - & - & - & - & - & - & $S_{60\mathrm{\mu m}}> 4
\mathrm{Jy}$\\
\multirow{2}{*}{Condon 2002} & \multirow{2}{*}{~~~~-} & \multirow{2}{*}{1897} & \multirow{2}{*}{8.2} & \multirow{2}{*}{10.8} & \multirow{2}{*}{287} & \multirow{2}{*}{42.2} & \multirow{2}{*}{51.6} & \multirow{2}{*}{1498} & \multirow{2}{*}{2.1} & \multirow{2}{*}{3.4} &
$S_{1.4\mathrm{GHz}}>2.5 \mathrm{mJy}$ \\
&  & & &  & &  &  & &  &  & and $m_{P}<14.5$ \\
\multirow{2}{*}{Drake 2003} & \multirow{2}{*}{~~~~-} & \multirow{2}{*}{178} &  \multirow{2}{*}{55.1} & \multirow{2}{*}{60.7} & \multirow{2}{*}{-} & \multirow{2}{*}{-} & \multirow{2}{*}{-} & \multirow{2}{*}{-} & \multirow{2}{*}{-} & \multirow{2}{*}{-} & $S_{4.8\mathrm{GHz}}\gtrsim
16 \mathrm{mJy}$ \\
 &  & &   & & & & &&  &  & and $S_{60\mathrm{\mu m}}\gtrsim 0.1 \mathrm{Jy}$ \\
Omar 2005 & Eridanus group & 72 & 2.8 & 2.8 & 20 & 5.0 & 5.0 & 46 & 2.2 & 2.2 & \\
Niklas 1995 &  Virgo cluster & 37 & - & 16.2 & 2 & - & 0.0 & 35 & - & 17.1 & radio@4.8~GHz \\
Andersen 1995 & Cluster \& group (poor) & 23 & 8.7 & 21.7 & - & - & - & - & - & - & \\
 & Cluster \& group (rich) & 20 & 15.0 & 25.0 & - & - & - & - & - & - & \\
Miller 2001 & Clusters $0<r<1$ Mpc & 120 & 28.3 & 37.5 & 54 & 46.3 & 53.7 & 53 & 3.8 & 17.0 &
$S_{1.4\mathrm{GHz}}>10 \mathrm{mJy}$ \\
 & Clusters $1<r<2$ Mpc & 96 & 21.9 & 29.2 & 23 & 60.9 & 73.9 & 50 & 6.0 & 10.0 & $S_{1.4\mathrm{GHz}}>10
\mathrm{mJy}$ \\
 & Clusters $2<r<3$ Mpc & 94 & 6.4 & 12.8 & 19 & 26.3 & 31.6 & 47 & 0.0 & 4.3 & $S_{1.4\mathrm{GHz}}>10
\mathrm{mJy}$ \\
Reddy 2004 & X-ray clusters & 114 & 13.2 & 19.3 & 33 & 30.3 & 45.5 & 81 & 6.2 & 8.6 & $L_{60\mathrm{\mu
m}}> 8.92 \mathrm{L_{\odot}}$ \\
 & X-ray clusters core & 33 & 24.2 & 39.4 & 15 & 40.0 & 66.7 & 18 & 11.1 & 16.7 & $L_{60\mathrm{\mu
m}}> 8.92 \mathrm{L_{\odot}}$ \\
 & X-ray clusters ring & 81 & 8.6 & 11.1 & 18 & 22.2 & 27.8 & 63 & 4.8 & 6.3 & $L_{60\mathrm{\mu
m}}> 8.92 \mathrm{L_{\odot}}$ \\
\hline
\end{tabular}
\begin{list}{}{}
\item[$^{\rm 1}$] The percentages are computed over the number of galaxies for each morphological subsample.
\item[$^{\rm 2}$] Number of galaxies in the total samples or the morphological subsamples.
\item[$^{\rm 3}$] F3: factor 3 radio excess; F5: factor 5 radio excess. Figures given in percentages.
\item[$^{\rm 4}$] All percentages for the fraction of radio-excess galaxies are upper limits as explained in Sect.~\ref{sec:error}.
\end{list}
\end{center}
\end{table*}

\begin{table*}
\caption{Rate of AGN candidates  in the literature from FIR colour.}
\begin{center}
\label{tabla:color-rate}
\begin{tabular}{l l c c }
\hline \hline
\multicolumn{1}{c}{} & \multicolumn{1}{c}{} & \multicolumn{2}{c}{Rate of AGN-candidates$^{\rm 1}$}  \\
Sample  &  Environment  & Total  & 2~Jy cutoff \\
\hline
AMIGA & Isolated & 28.4 & 14.3 \\
Condon 1991 & ~~~~- & 21.8 & 12.4  \\
Yun 2001 & ~~~~- & 15.9 & 15.9  \\
Condon 2002 & ~~~~- & 13.3 & -  \\
Drake 2003 & ~~~~- & 45.0 & 22.6  \\
Andersen 1995 & Cluster \& group (poor) & 56.5 & -  \\
              & Cluster \& group (rich) & 75.0 & -  \\
Reddy 2004 & X-ray clusters & 59.0 & 20.0  \\
\hline
\end{tabular}
\begin{list}{}{}
\item[$^{\rm 1}$] Figures given in percentages.
\end{list}
\end{center}
\end{table*}

We notice that a detailed comparison of these papers is difficult for 
a number of reasons that we list below. The selection criteria of 
the samples differ significantly, introducing different biases 
depending on whether the selection was performed at optical 
wavelengths or with a FIR or radio cut selection criterion. Normalisation of the
percentages is also a delicate issue since, in some cases, the values are divided
by the total number of galaxies with available data, but in others only
detections are used. In the cases where AGN selection was performed using the 
radio-FIR 
correlation, the wavelength of the radio-emission is usually 1.4~GHz but in some
cases it is 4.8~GHz. The radio excess cutoff varies for different papers,
the most used factors are 3 times or 5 times 
above the radio-FIR correlation. Finally, the used radio-FIR correlation is, in some
cases, the one obtained for the sample analysed in the paper, but
in others the one for a different reference sample is used, e.g., 
\citet{Omar2005} used \citet{Yun2001} correlation.

We have obtained the figures given in Table~\ref{tabla:exceso-rate} 
by reanalysing the data given in the papers  (e.g. most of
them had not performed an analysis by morphological type) or provided by the
authors, in an attempt to homogenise the statistics. We have defined the radio excess
in two ways: a factor 3 or a factor 5 above the radio-FIR correlation for
our sample of isolated galaxies, corresponding to  \mbox{$q < 1.88$} or \mbox{$q < 1.66$},
respectively. The percentages are calculated with respect to the total number of
galaxies that can be classified as explained in Sect.~\ref{sec:radio-excess}. 
The rates for different
morphological types are not calculated with respect to the total number of
galaxies but normalised for the corresponding subsamples of early and late-type
galaxies. Comparison of the AGN candidate rate based on the FIR colour selection
was possible for some of the references given above
\citep{Condon1991b,Yun2001,Drake2003,Andersen1995,Reddy2004}. In these  studies,  
$S_{25\mathrm{\mu m}}$ and $S_{60\mathrm{\mu m}}$ fluxes or $\alpha_{25,60}$
(spectral index between $S_{25\mathrm{\mu m}}$ and $S_{60\mathrm{\mu m}}$) were
given, and we selected as AGN candidates those with $S_{25\mathrm{\mu
m}}/S_{60\mathrm{\mu m}} \geq 0.18$  (see Sect.~\ref{sec:IRAS-colours}) equivalent
to $\alpha_{25,60} < 1.958$. We have to be careful when comparing these numbers,
since, e.g., the sample of  \citet{Yun2001} has been selected to have  $S_{60\mathrm{\mu
m}} \geq 2~\mathrm{Jy}$. This will bias the result since the AGN candidate rate
depends directly on this variable. However, we think that this cut has a
significant advantage since much of the AGN candidates have low IRAS fluxes
(Fig.~\ref{fig:iras_selection}) near the detection limit. Therefore,
the error is very high in comparison to the measurements,
hence confidence in the selection of AGN candidates gets
lower. For this reason, we have both provided the statistics using the
original data in each paper and  applying  the $S_{60\mathrm{\mu m}} \geq
2~\mathrm{Jy}$ cut to all samples when it was possible, including ours (see
Table~\ref{tabla:color-rate}).

\subsubsection{``Field'' environments}

As indicated above some studies of radio-excess galaxies 
 were performed for samples usually referred to as
``field'' in the sense that they were not selected in any particular way
with respect to their environment. However, it is found that, for instance, many cD
galaxies belong to these ``field'' samples. We comment below on these studies.

\citet{Condon1991b}  studied a sample of  IRAS sources applying a high radio flux
cutoff at 4.75~GHz and with no definition of the environment (e.g., several
pairs are included in their sample). They also apply a FIR cutoff but not as
strict as  the radio one. Using their data in Table~2 we have calculated the
ratios of radio-excess listed in our Table~\ref{tabla:exceso-rate} and that of
AGN candidates based on $\alpha_{25,60}$ as given in
Table~\ref{tabla:color-rate}. When the 2~Jy cut is applied to their sample 
the percentage of galaxies with $\alpha_{25,60}< 1.958$ decreases, as
expected. The radio flux cut leads also to a bias since it will  increase the
ratio of radio-excess galaxies with respect to  other samples listed in
Table~\ref{tabla:exceso-rate},  selected  without such a cutoff or using a
higher FIR cutoff, the latter obviously producing the opposite bias. The
separation of the data by morphological types clearly shows a higher rate of
radio excess for early-type galaxies.

\begin{figure}
\centering
\resizebox{9.5cm}{9.5cm}{\includegraphics{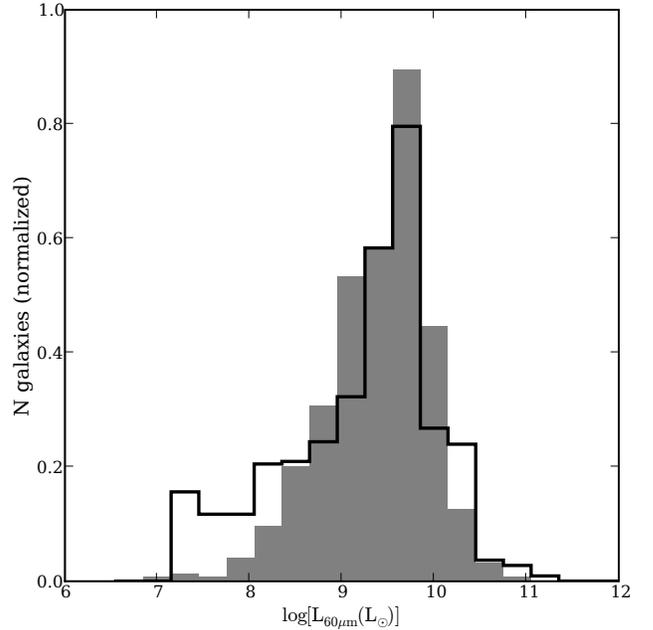}}
   \caption{Histogram of $L_{60 \mathrm{\mu m}}$ luminosity for our complete 
subsample ($n=710$; black line) and the \citet{Yun2001} sample ($n=1248$; grey bars).
           }
      \label{fig:l60yun}
\end{figure}

\citet{Yun2001} studied in detail an IR flux-limited complete sample with no
selection concerning the environment. Their fit to the radio-FIR correlation is
identical to ours within the errors since both samples are dominated by star
formation activity, with very few galaxies deviating from the correlation, hence
leading to very similar q-parameters, 2.34 in their sample versus 2.36 in ours.
Their ratio of radio-excess galaxies being very low, is still a factor 3 larger
than ours. Two main differences are found between both samples. While our sample is
optically selected, \citeauthor{Yun2001} selected their sample in the FIR. Applying the same
FIR flux cutoff to our complete subsample leads to a decrease of the ratio of
radio-excess galaxies to 0\%. On the other hand, galaxies in their sample
lie in higher density environments than ours: among the 23 radio excess galaxies
in their sample 2 are giant E/S0 in clusters and 16 galaxies are
interacting in pairs or higher ranked environments. Furthermore, a comparison of
the  $\log L_{60 \mathrm{\mu m}}(\mathrm{L_{\odot}})$ distribution for both
samples (Fig.~\ref{fig:l60yun}) shows a higher ratio of galaxies at the low
luminosity end for the AMIGA sample, with a difference in the mean of 0.31.  Since
the FIR luminosity is a variable widely known to be driven by interaction this
supports further the higher density of the environment traced by the sample in
\citet{Yun2001}. Unfortunately, no morphological types are given for their full
sample and we could not check the percentages for early and late-type galaxies.
They kindly provided the $S_{25\mathrm{\mu m}}$ and $S_{60\mathrm{\mu m}}$
fluxes and we used them to derive the statistics of AGN
candidates based on IRAS colours for their sample
(Table~\ref{tabla:color-rate}).

\citet{Corbett2002}  studied a sample of objects with  IRAS data to
perform a comparison of the multiwavelength properties of a sample of AGN with
and without compact radio cores and a matched sample of galaxies without an AGN.
They imposed a FIR flux limit higher than \citet{Yun2001}. As a consequence, the 
\citeauthor{Corbett2002} selected galaxies have FIR luminosities greater than
$10^{10.5}~\mathrm{L_{\odot}}$ which, as they indicate, results in an enhanced
fraction of interacting galaxies in the sample. They estimate that 40\% of their
sample is obviously involved in either tidal interactions or are major mergers,
and 20\% are apparently non-interacting. 
Using the data in Table~2 of \citet{Corbett2002}, we obtained their ratio
of radio-excess galaxies. We found a slightly higher value of radio-excess galaxies
than for the sample of \citet{Yun2001}; the \citet{Corbett2002} data yielded 3 times 
our value, but is still only a few percent. 
This low value is, as they indicate, most likely due to their
high FIR cutoff, artificially lowering the rate of radio-excess galaxies.

\citet{Condon2002} studied an Uppsala General Catalogue of Galaxies (UGC) sample 
composed of  sources detected in the
NVSS and covered by the IRAS survey, using a radio-flux cutoff and no selection in
terms of the environment. Using the data in their Table~1, we have calculated
their rate of radio-excess galaxies, both total and separated by morphological types,
as well as the percentage of  AGN candidates with \mbox{$\alpha_{25,60} <
1.958$}. We could not apply a cut at $S_{60\mathrm{\mu m}} \geq 2~\mathrm{Jy}$
since  $S_{60\mathrm{\mu m}}$ was not available. We note that the radio
cutoff for this sample is lower than in \citet{Condon1991b}, and the ratio of
radio-excess galaxies is also lower.  As in \citet{Condon1991b}  a higher rate of
radio excess is found for early-type galaxies.

\citet{Drake2003} used a sample selected by cross-correlating the IRAS Faint
Source Catalogue with the 5~GHz Parkes-MIT-NRAO catalogue. We  derived the ratio
of AGN candidates with $S_{25\mathrm{\mu m}}$/$S_{60\mathrm{\mu m}} \geq 0.18$
using the data in their Table~1 and list the value with and without the 2~Jy cut
in Table~\ref{tabla:color-rate}. They find a radio-excess rate of $\approx
55\%$, using a factor 5 cutoff at 4.8~GHz with respect to the  mean of the $u$
parameter in \citet{Condon1995} ($u \equiv \log(S_{60\mathrm{\mu
m}}/S_{4.8~GHz})$; $<u> = 2.5$). They sample lower IRAS flux densities than
previous authors and thus find a larger fraction of objects with intermediate radio
fluxes. In a
later paper \citep{Drake2004a} the host galaxies of radio-excess sources are
studied, finding that only 23\% of them are isolated and with no evidence of
disturbance. They also determine the morphology of the hosts, obtaining that
elliptical galaxies show a large range of radio excess, while the disk systems
all have only a moderate radio excess.

\subsubsection{Dense environments}

\citet{Niklas1995} analysed the radio-FIR correlation for Virgo cluster galaxies
at 4.8 and 10~GHz. Inspection of their Fig.~4 indicates that they used a cutoff of a factor
3 to select radio-excess galaxies with respect to their radio-FIR
correlation. We have used the data in their paper to obtain the remaining values
given in Table~\ref{tabla:exceso-rate}.

\citet{Andersen1995} compared the FIR-radio correlation for spirals in rich and
poor cluster and group environments. Using the data in their Table~4 and~5 we
calculated the radio-excess ratio in their subsample.

\citet{Miller2001} considered a sample of nearby Abell clusters galaxies. 
They kindly provided the data used in their paper which we have
used to estimate the ratio of radio-excess galaxies in their sample for 
different distances from the cluster center, as we list in Table~\ref{tabla:exceso-rate}. 
They considered 3 subsamples: one including galaxies in the inner Mpc of clusters, one 
in the 1~-~2~Mpc range and the last one at the
periphery of clusters, between 2 and 3~Mpc from the cluster core in
projection. The last subsample is likely to be very similar to field-like environments.
They find a considerable gradient in the ratio of
radio-excess galaxies from the outer parts to the centre of the cluster (see Table~\ref{tabla:exceso-rate}).
They suggest that this is probably
a result of the presence of centrally concentrated radio-luminous elliptical galaxies. No
data on the morphological types are given in the paper. In order to
investigate whether the same trend holds for spiral galaxies, we compiled the
morphologies from NED. For the low density parts of Abell clusters we find a
very low ratio of spirals with radio-excess and a much higher one for early types. 
For the inner 2 Mpc the ratio of radio-excess galaxies increases by a
factor 2-3 both for early and late-type galaxies.

\citet{Reddy2004} studied a sample of cluster galaxies located in the seven
nearest clusters with prominent X-ray emission using a cut in the FIR
luminosities. They use $q < 1.64$ as cutoff for radio-excess galaxies, very
similar to our value (1.66) but for coherence we recalculated the ratios referred to
our value. We also estimated the ratios for early and late-type galaxies using
the data in their Table~2. Their ratios are similar to the values in other
clusters, except for those cases where a radio flux cutoff was applied, in which case the
radio-excess ratios are higher due  to selection effects. We provide also the
ratios for galaxies with projected cluster-centric distances lower than 0.5~Mpc
and between 0.5 and 1.5~Mpc, defined by them as core and ring respectively.
These data are not given in their paper and were kindly provided by the authors. We
notice that, when divided into early and late-type galaxies, both early and late
types show a significant increase in the ratio of radio-excess for smaller
radii. This result cannot be directly compared with \citet{Miller2001}
since the area defined as ring by \citet{Reddy2004}  is still in the inner
area of the \citeauthor{Miller2001} sample. We used the $S_{25\mathrm{\mu m}}$ and
$S_{60\mathrm{\mu m}}$ fluxes provided  by them  to derive the statistics
of AGN candidates based on IRAS colours for their sample
(Table~\ref{tabla:color-rate}).

\citet{Omar2005} studied the radio-FIR correlation for the Eridanus group, which
is not dynamically relaxed. They consider the 72 galaxies detected in the IRAS
survey. Using the data in their Fig.~3 and Table~2 we have calculated the radio
excess for the total sample and divided into morphological types.

\subsubsection{Radio-excess, environment and density-morphology relation}

A comparison with  the above discussed samples from the literature shows that
the AMIGA sample has the lowest ratio of radio-excess galaxies, both 
globally and separated into early and late types (Table~\ref{tabla:exceso-rate}).
Higher ratios of radio excess galaxies are found  in
denser but still poor environments as the outer parts of clusters with prominent
X-ray emission \citep{Reddy2004}, outer parts of Abell clusters 
\citep{Miller2001}, poor cluster and group environment
\citep{Andersen1995}, Virgo cluster \citep{Niklas1995} and
Eridanus group  \citep{Omar2005}. The  values in the outermost, field-like
sample of \citeauthor{Miller2001} are similar to the ones in \citet{Condon2002},
 suggesting that the outer parts of clusters are as efficient as the ``field''
environment to trigger radio emission in active galaxies. For the densest
environments an even higher ratio of radio-excess is found, as in the
core sample of \citeauthor{Reddy2004}, the inner area of Abell clusters
\citep{Miller2001} or rich environments \citep{Andersen1995}. It
is interesting to note that the \citeauthor{Drake2003} sample is dominated by
one-to-one interactions, i.e., higher local density, and its radio-excess rate
is the highest among all. Although the radio-flux cutoff used in their sample is
high, hence producing an artificial increase in radio-excess galaxies, the
result might also reflect that one-to-one interactions are more efficient in
inducing radio activity in galaxies than the larger scale environment. This
could be explained by the  fuelling of gas towards the centre of galaxies in
interacting pairs, feeding the AGN and leading to the radio-excess regime.

The same result is  found using FIR colours, independently whether the 2~Jy cut
is applied or not: we find that the rate of AGN candidates based on IRAS
colours increases from the most isolated environments to the densest ones. We
notice that when the 2 Jy  cut is applied, although the trend is kept the gradient is
significantly reduced, confirming the limited reliability of the method. 

Finally, we have studied several subsets of the AMIGA sample, 
taking into account the refinement of the CIG with respect to the environment
that our group carried out. In our
reevaluation of the optical morphologies of the CIG galaxies we identified 32
objects candidates to be suffering interactions based on evidence for
asymmetries/distortions that might be of tidal origin \citep{Sulentic2006}. 
None of  these galaxies show a radio excess above a cutoff of 3. A further
revision of the sample showed that  150 additional galaxies lie in environments that
could affect their evolution based on the value of their local number density of
neighbours and the tidal forces at play \citep{Verley2007b}. These galaxies are 
uniformly distributed with respect to the radio-FIR correlation suggesting that
weak interactions in a low density environment do not significantly  affect
the radio activity of galaxies.

The higher ratio of radio-excess galaxies in denser environments has often been
explained as due to the density-morphology relation. In all the
samples studied in this paper,
a larger ratio of radio-excess galaxies is found for early types, which is not
surprising since usually the typical host galaxies of an AGN  with
powerful radio continuum emission  are massive ellipticals. The higher
abundance of early-type galaxies in denser environments could by itself justify
the higher ratio of radio active galaxies. We think however that although it
might explain in part the results, the environment also has to play a prominent 
role in triggering the radio
activity. On one hand the ratio of early-type galaxies with
radio-excess in our sample, higher than for late-type galaxies,  is still
much  lower than for all other environments (only 4.8\% of the early-type
galaxies have a radio excess above a factor 5 cutoff). This result might be
explained by the low luminosities of the AMIGA early-type population relative to
the AMIGA spiral population and to early-type populations found  in most surveys
 \citep{Sulentic2006}. This shows that the low ratio of radio-excess
galaxies in our sample can not be only due to the small percentage of early-type galaxies 
(12\% in the complete subsample, see
Table~\ref{tabla:exceso}), as expected for a low-density environment. On the
other hand, we find evidence that spiral galaxies also increase their rate of
radio activity with environment. The ratios for the spirals in our sample of
isolated galaxies are clearly lower than for samples in denser environments. 

\section{Conclusions}
\label{sec:conclusions}

In this paper, we have studied the rate of AGN candidates in a well-defined 
complete sample of isolated galaxies as part of the AMIGA project.
We have focussed on two methods that make use of the 
radio (NVSS) and FIR (IRAS) data for our sample, and complemented
them with additional data found in the literature. Our main
results for the AMIGA sample are:
\begin{itemize}
\item Our sample is mostly radio quiet, with most galaxies ($98.6 \%$)
having radio powers lower than  $10^{23}~\mathrm{W~Hz}^{-1}$, consistent
with the high ratio of late-type galaxies in our sample.
\item We have selected  radio-excess candidate galaxies above
the radio-FIR correlation
for our complete subsample, and revised the results using
FIRST data to exclude back/foreground sources.
We find less than 0.8\% radio-excess galaxies with an excess of a
factor 5 and less than 4.0\% for a lower excess of a factor 3.
\item Using the IRAS flux ratio   $ S_{25\mathrm{\mu m}}/S_{60\mathrm{\mu m}}$
 to select AGN candidates
we find a frequency of AGN candidates of $\backsim 28\%$ with a lower 
limit of $\backsim 7\%$.
\item From NED and the V\'eron-Cetty catalogues, we found
$n=29$ AGN candidates (including LINERs and NLAGN).
\item The final catalogue contains a total of 89 AGN
candidates. This catalogue is available in electronic form at the
CDS\footnote{ftp://cdsarc.u-strasbg.fr/} and at the AMIGA web
page\footnote{http://www.iaa.csic.es/AMIGA.html}.
\end{itemize}

We have compared our results with those found in the 
literature and interpreted them, taking into account
that our sample was selected using optical
criteria  and that we used a well-defined
criterion of isolation.  We conclude that:
\begin{itemize}
\item The AMIGA sample has the lowest ratio of AGN candidates, both
globally and considering early-type and late-type galaxies separately.
\item Field galaxies as well as galaxies in poor cluster and group environments (e.g., outer parts 
of clusters) show intermediate values, although the numbers
are only illustrative as numerous selection effects affect the conclusions.
The outer parts of clusters seem as efficient as
field/poor environments to trigger radio emission in AGN.
\item The highest rates of AGN candidates are found in 
the central parts of clusters, but also in pair/merger dominated
samples.
\item For all environments,  a higher ratio of radio excess is found 
for early-type galaxies, as can be expected since massive
ellipticals are the usual hosts for powerful radio jets.
\item Both elliptical and spiral galaxies increase their radio excess activity for
denser environments. This increment supports that the density-morphology
relation is not the only explanation for the enhancement in
AGN frequency in denser environments, i.e., nuclear 
activity is not only associated with dense environments
due to its higher content in ellipticals, but is directly triggered
by it. 
\end{itemize}

Hence, the environment seems to play a crucial role in the development of
nuclear activity both at large scales and in strong one-to-one
interactions. 

Finally, we notice that the AMIGA sample appears to represent the most 
nurture-free population of luminous early-type galaxies
as confirmed by their lack of radio excess above the FIR-radio correlation.
The catalogue presented here can be used as 
a baseline for forthcoming studies about the relation between environment
and nuclear
activity. We plan to complement it
with optical spectroscopy data.

\begin{acknowledgements}
We would like to warmly thank everybody who provided data for this study. 
The authors would also like to acknowledge the anonymous referee for his/her 
suggestions to improve the paper.

LVM, UL, SL, JS and JSM are partially supported by DGI Grant
AYA 2005-07516-C02-01 and Junta de Andaluc\'{\i}a (Spain). SL was partially supported
by an Averroes fellowship contract from the Junta de Andaluc\'{\i}a. UL is supported by
a Ramon y Cajal fellowship contract and by the DGI Grant ESP2003-00915. 
JSM is supported by a fellowship from the
Secretar\'{\i}a de Estado de Educaci\'on y Universidades.

This research has made use of the NASA/IPAC Extragalactic Database (NED)
which is operated by the Jet Propulsion Laboratory, California Institute
of Technology, under contract with the National Aeronautics and Space
Administration.
\end{acknowledgements}

\bibliographystyle{aa}
\bibliography{databasesol}

\end{document}